\documentclass[12]{article}
\usepackage{amssymb}
\usepackage{graphics}
\begin{document}
\mbox{}\\[2cm]
\begin{center}
{\LARGE \bf Aperiodic packings of clusters obtained by projection\\[5mm] }
NICOLAE COTFAS\\[3mm]
Faculty of Physics, University of Bucharest, PO Box 76-54, Post Office 76, Bucharest, Romania\\[3mm]
E-mail: ncotfas@yahoo.com  \quad Homepage at \verb#http://fpcm5.fizica.unibuc.ro/~ncotfas# \\[1cm]
\end{center}
{\bf Abstract.}
Atomic-resolution electron microscope images show that a quasicrystal is a 
quasiperiodic packing of clusters. The outer atomic shells of multi-shell
clusters occuring in quasicrystals are highly symmetric and rather robust, 
but some structural disorder may be present around the core.
The models describing quasicrystal structure are quasiperiodic lattices 
containing two or more unit cells decorated by atomic clusters. 
We show that a modified version of the strip projection method may become an
alternate way to obtain useful models.\\[5mm]

\section{Introduction}

It is known that all quasicrystals found so far are composed of building units, 
usually called atomic clusters \cite{hbs,yt,ayp,ws,gpqk,jb,rg,hkch}.
An atomic cluster is a set of close atoms, generally, distributed on fully occupied 
high symmetry shells, which are found at a high frequency in quasicrystal \cite{gpqk,kp}.
Atomic-resolution electron microscope images offer direct information on the atomic clusters 
occuring in quasicrystal and cluster packing.
The outer atomic shells of multi-shell clusters occuring in quasicrystals are highly 
symmetric and rather robust, but some structural disorder may be present around the core
\cite{ayp}.

Any finite set admitting a group $G$ as a symmetry group is a union of orbits of $G$,
and is usually called a $G$-cluster. A mathematical algorithm 
for generating quasiperiodic point sets by starting from $G$-clusters 
was proposed by author in collaboration with Verger-Gaugry several
years ago \cite{c97}. It is based on strip projection method 
and is a direct generalization of the algorithm presented by Katz and Duneau
in \cite{k86}. The model used for the icosahedral quasicrystals in \cite{k86} starts
from the one-shell icosahedral cluster $\mathcal{C}$ formed by the vertices of a regular
icosahedron. The physical space is embedded into the superspace $\mathbb{R}^6$ such 
that the orthogonal projections on the physical space of the points
\[ (\pm 1,0,0,0,0,0),\, (0,\pm 1,0,0,0,0),\, ...,\, (0,0,0,0,\pm 1,0),\, 
(0,0,0,0,0,\pm 1) \]
are the vertices of a regular icosahedron.
 
In our direct generalization, we consider only $G$-clusters invariant under inversion.
If our starting $G$-cluster $\mathcal{C}$ has $2k$ points then we embed the physical
space into the superspace $\mathbb{R}^k$ in such a way that 
$\mathcal{C}$ is the orthogonal projection on the physical space of the subset
\[ \{ (\pm 1,0,0,...,0),\, (0,\pm 1,0,0,...,0),\, ...,\, (0,0,...,0,\pm 1,0),\, 
(0,0,...,0,\pm 1)\} \]
of $\mathbb{R}^k$ containing $2k$ points. Strip projection allows to obtain a pattern $\mathcal{P}$ such that  for each point $x\in \mathcal{P}$ the neighbours of $x$ belong to the translated copy
$x+\mathcal{C}$ of $\mathcal{C}$. This means that $\mathcal{P}$ is a quasiperiodic packing
of interpenetrating partially occupied copies of $\mathcal{C}$. In the case of a multi-shell cluster our algorithm uses a superspace of rather high dimension and the occupation of 
clusters occurring in the obtained pattern is extremely low.

Quasicrystals are materials with perfect long-range order, but with no three-dimensional translational periodicity. The description of atomic structure of quasicrystals is 
a highly non-trivial task, and we are still far away from a satisfactory quasicrystal
structure solution \cite{ws}. The quasicrystal structure is usually described by using a 
quasiperiodic lattice containing two or more unit cells decorated with atoms. 
This description is purely mathematical and does not provide any physical insight 
on why the atoms should favour such a complicated structure \cite{ayp}. Our aim is to present
a modified version of the strip projection method which leads to useful models and might be
justified from a physical point of view.

\section{Packings of $G$-clusters defined in terms of the strip projection method}

Let $G=D_n$ be one of the dihedral groups $D_8$, $D_{10}$, $D_{12}$.
The group $G$ can be defined as
\begin{equation}
 G=\langle \ a,\ b\  |\ \ a^n=b^2=(ab)^2=e\ \rangle 
\end{equation}
and the formulae
\begin{equation}\begin{array}{l}
 a(\alpha ,\beta )=
\left(\alpha \, \cos \frac{2\pi }{n}-\beta \, \sin \frac{2\pi }{n}, \
      \alpha \, \sin \frac{2\pi }{n}+\beta \, \cos \frac{2\pi }{n} \right)\\
b(\alpha, \beta )=(\alpha ,-\beta )
\end{array}
\end{equation}
define an $\mathbb{R}$-irreducible representation in $\mathbb{R}^2$. 
Let 
\begin{equation} 
\mathcal{C}=\{ v_1,\, v_2,\, ...,\, v_k,\, -v_1,\, -v_2,\, ...,\, -v_k \}
\end{equation} 
where $v_1=(v_{11},v_{21})$, $v_2=(v_{12},v_{22})$,..., $v_k=(v_{1k},v_{2k})$,
be a fixed $G$-cluster symmetric with respect to the origin.
From our general theory \cite{c97} (a direct verification is also possible) it
follows that the vectors 
\begin{equation} 
w_1=(v_{11},v_{12},...,v_{1k})\qquad {\rm and}\qquad 
 w_2=(v_{21},v_{22},...,v_{2k})
\end{equation}
from $\mathbb{R}^{k}$ are orthogonal and have the same norm
\begin{equation} 

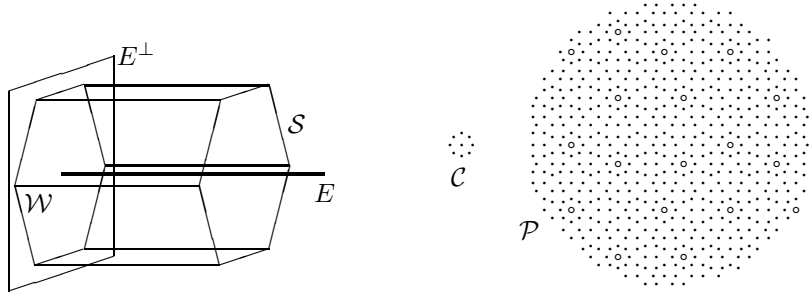
\caption{{\it Left:} The strip $\mathcal{S}=E+[-1/2,1/2]^3$ and the 
window $\mathcal{W}=\pi ^\perp ([-1/2,1/2]^3)$ in the case of a
one-dimensional physical space $E$ embedded into the three-dimensional superspace $\mathbb{R}^3$.
{\it Right:} The one-shell $C_8$-cluster $\mathcal{C}=C_8(1,0)$  
and a fragment of the set $\mathcal{P}$ defined by using 
this cluster and strip projection method in a four-dimensional superspace.
The nearest neighbours of any point $p\in \mathcal{P}$ belong to $p +\mathcal{C}$, 
which is a copy of $\mathcal{C}$ with the centre at point $p$. The centres of fully occupied 
clusters are indicated by $\circ $.} 		   
\end{figure}

\begin{figure}
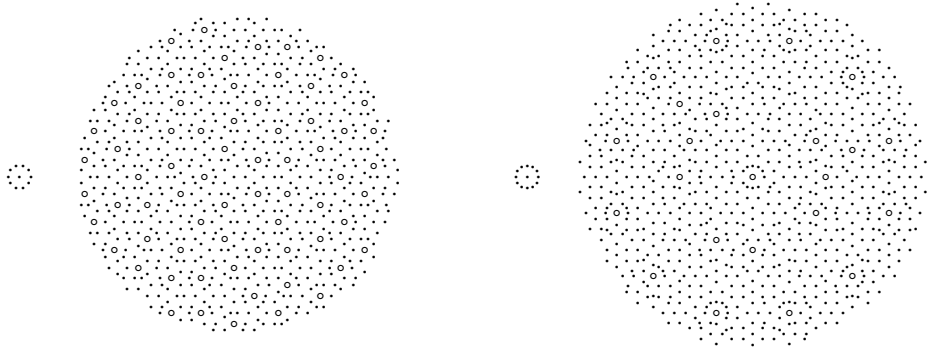

		   
\caption{{\it Left:} A fragment of the set defined by using the one-shell  
$C_{10}$-cluster $\mathcal{C}=C_{10}(1,0)$ and strip projection method in a five-dimensional superspace. This pattern does not contain fully occupied clusters, and its points represent 
the vertices of a Penrose tiling. The centres of the clusters with occupation greater than 
50\% are indicated by $\circ $. {\it Right:} A fragment of the set defined by using the one-shell  
$C_{12}$-cluster $\mathcal{C}=C_{12}(1,0)$ and strip projection method in a six-dimensional superspace. The centres of the clusters with occupation greater than 
50\% are indicated by $\circ $.} 
\end{figure}

\begin{figure}[h]
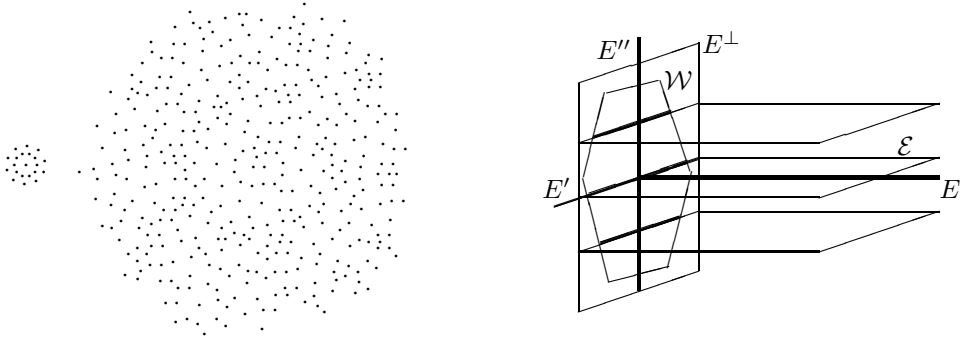


\caption{{\it Left:} A fragment of the set $\mathcal{P}$ defined by using the two-shell cluster 
$\mathcal{C}=C_{10}(1,0)\cup C_{10}(1.1,1.3)$ and strip projection method in a ten-dimensional superspace. The nearest and the next-to-nearest neighbours of any point $p\in \mathcal{P}$ 
belong to $p +\mathcal{C}$, but the occupation of the copies of $\mathcal{C}$ occurring in 
this pattern is extremely low.  
{\it Right:} The superspace decompositions $\mathbb{R}^k=E\oplus E^\perp=
E\oplus E'\oplus E''={\mathcal E}\oplus E''$. The points of $\mathbb{Z}^k$ lie 
in a family of subspaces parallel to $\mathcal{E}=E\oplus E'$.}
\end{figure}

We identify the physical space with the two-dimensional subspace
\begin{equation} 
E=\{ \ \alpha w_1+\beta w_2 \ | \ \alpha,\, \beta \in \mathbb{R}\ \} 
\end{equation}
of the superspace $\mathbb{R}^k$ and denote by $E^\perp $ the 
orthogonal complement (figure 1)
\begin{equation} E^\perp =\{ \ x\in \mathbb{R}^k\ |\ 
\langle x,y\rangle =0\ {\rm for\ all}\ y\in E\ \}. \end{equation}
The orthogonal projection on $E$ of a vector $x\in \mathbb{R}^k$ is the vector
\begin{equation} \label{pi}
\pi \, x= \left\langle x,\frac{w_1}{\kappa }\right\rangle\frac{w_1}{\kappa }+
             \left\langle x,\frac{w_2}{\kappa }\right\rangle\frac{w_2}{\kappa} 
\end{equation}
where $\kappa =||w_1||=||w_2||$, and the orthogonal projector corresponding 
to $E^\perp $ is
\begin{equation} 
\pi ^\perp :\mathbb{R}^{k}\longrightarrow E^\perp \qquad
\pi ^\perp x=x-\pi \, x. 
\end{equation}
We describe $E$ by using the orthogonal basis 
$\{ \kappa ^{-2}w_1,\, \kappa ^{-2}w_2\}$, and therefore, the expression in coordinates of $\pi $ is
\begin{equation} 
\pi : \mathbb{R}^{k}\longrightarrow \mathbb{R}^2\qquad 
\pi x=(\langle x,w_1\rangle , \langle x,w_2\rangle ). 
\end{equation}

The set defined in terms of the {\it strip projection method} \cite{k86,e86,c06a,c06b}
\begin{equation}
\mathcal{P}=\pi (\mathcal{S}\cap \mathbb{Z}^{k})=
\{ \ \pi x\ | \ \ x\in \mathcal{S}\cap \mathbb{Z}^{k}\ \} 
\end{equation}
by using the strip (figure 1)
\begin{equation} 
\mathcal{S}=\Lambda +E=\{ \ x+y\ | \ x\in \Lambda , \ y\in E \ \} 
\end{equation}
generated by shifting along $E$ the unit hypercube
\begin{equation}
\Lambda =\left\{ (x_1,x_2,...,x_{k})\ \left|\ -\frac{1}{2}\leq x_i\leq \frac{1}{2}\ \
{\rm for\ all\ } i\in \{ 1,2,..., k\}\ \right. \right\} 
\end{equation}
is a packing of partially occupied copies of $\mathcal{C}$.
The set of arithmetic neighbours of a point $x\in \mathbb{Z}^k$  is
\begin{equation} 
\mathcal{A}(x)=\{ \ x+e_1,\, x+e_2,\, ...,\, x+e_k,\, x-e_1,\, x-e_2,\, ...,\, x-e_k\ \}
\end{equation} 
where $e_1=(1,0,...,0)$, $e_2=(0,1,0,...,0)$, ..., $e_k=(0,...,0,1)$ are the vectors 
of the canonical basis of $\mathbb{R}^k$, and the set of neighbours of a point 
$\pi x\in \mathcal{P}$ is $\pi (\mathcal{S}\cap \mathcal{A}(x))$. Since
\begin{equation}
\pi e_i=(\langle e_i,w_1\rangle , \langle e_i, w_2\rangle )=(v_{1i},v_{2i})=v_i
\end{equation}
we get
\begin{equation}
 \pi (\mathcal{S}\cap \mathcal{A}(x))
 \subseteq \{ \pi x+v_1,\, \pi x+v_2\, ...,\, \pi x+v_{k},\pi x-v_1,\, 
\pi x-v_2\, ...,\, \pi x-v_{k}\}=\pi x+\mathcal{C} 
\end{equation}
that is, the neighbours of any point $\pi x\!\in \!\mathcal{P}$ belong to the translated copy 
$\pi x\!+\!\mathcal{C}$ of $\mathcal{C} $.
A larger class of quasiperiodic patterns can be obtained by translating the strip $S$. 
The set
\begin{equation} \label{omega} 
\mathcal{P}=\pi ((t+\mathcal{S})\cap \mathbb{Z}^{k})=
\{ \ \pi x\ | \ \ x\!-\!t\in \mathcal{S}\ \ {\rm and}\ \ x\in \mathbb{Z}^{k}\ \} 
\end{equation}
is a packings of partially occupied copies of $\mathcal{C}$, for any $t\!\in \!\mathbb{R}^k$.  

In the right hand side of figure 1 we present of a fragment a pattern $\mathcal{P}$ defined by
starting from the $C_8$-cluster
\[
\mathcal{C}\!=\!C_8(1,0)\!=\!\left\{ \pm (1,0),
\pm \left( \frac{1}{\sqrt{2}},\frac{1}{\sqrt{2}}\right), \pm (0,1),
\pm \left( \frac{-1}{\sqrt{2}},\frac{1}{\sqrt{2}}\right)\right\} .
\]
In this case, the dimension of the superspace is $k=4$,
\[
E=\left\{ \left. \alpha \left( 1,\frac{1}{\sqrt{2}},0,\frac{-1}{\sqrt{2}}\right)+
\beta \left( 0,\frac{1}{\sqrt{2}},1,\frac{1}{\sqrt{2}}\right) \ \right| \ \alpha , \beta \in \mathbb{R}\ \right\}
\]
${\rm dim}\, E^\perp ={\rm dim}\, E=2$, and one can remark that $\mathcal{P}$ contains fully
occupied clusters. The points of $\mathcal{P}$ are the vertices of an Ammann-Beenker tiling.

The set from the left hand side of figure 2, obtained by starting from the $C_{10}$-cluster 
$\mathcal{C}=C_{10}(1,0)$, is formed by the vertices of a Penrose tiling and does not contain 
fully occupied clusters. In this case the dimension of the superspace is $k=5$ and 
${\rm dim}\, E^\perp =3=1+{\rm dim}\, E$. If we start from the $C_{12}$-cluster $\mathcal{C}=C_{12}(1,0)$ 
then we obtain a set which does not contain fully occupied clusters (right hand side of figure 3).
In this case $k=6$ and ${\rm dim}\, E^\perp =4=2+{\rm dim}\, E$. 
The set from the left hand side of figure 5 is generated by starting from the two-shell 
$C_{10}$-cluster $\mathcal{C}=C_{10}(1,0)\cup C_{10}(1.1,1.3).$ 
The nearest and the next-to-nearest neighbours of any point $\omega $ belong to 
$\omega +\mathcal{C}$, but the occupation of the copies of $\mathcal{C}$ 
occurring in this pattern is extremely low. The dimension of the superspace used in this case is $k=10$
and ${\rm dim}\, E^\perp =8=6+{\rm dim}\, E$. These examples show that the occupation of the clusters occurring in our patterns diminish quickly when ${\rm dim}\, E^\perp -{\rm dim}\, E$ increases. 

One can prove \cite{c06b} that the space $E^\perp $ can be decomposed into a direct sum 
$E^\perp =E'\oplus E''$ such that $\mathbb{Z}^k$ is contained in a discrete family of
affine spaces parallel to $\mathcal{E}=E\oplus E'$ (see right hand side of figure 3). 
Only a finite number of these subspaces 
meet the window $\mathcal{W}=\pi ^\perp (\Lambda )$ corresponding to the strip $\mathcal{S}=\Lambda +E$. 
In the case when ${\rm dim}\, E^\perp ={\rm dim}\, E$ the space $E''$ is the null space $\{ 0\}$
and the set $\pi ^\perp (\mathcal{S}\cap \mathbb{Z}^k)$ is dense in the window
$\mathcal{W}=\pi ^\perp (\Lambda )$ corresponding to the strip $\mathcal{S}$.
This explains why we get fully occupied clusters in the case of $\mathcal{C}=D_8(1,0)$ and in 
the case of the model used by Katz and Duneau \cite{k86} and independently by Elser \cite{e86}
for icosahedral quasicrystals.

\section{Modified strip projection method}

The distances to the 'physical' space $E$ of the points of $\mathbb{Z}^k$ form a discrete 
sequence. In table 1 we present the smallest terms of this sequence in the cases presented
in figures 1 and 2. One can see why the occupation of clusters is decreasing when we pass 
from $C_8(1,0)$ to $C_{10}(1,0)$ and $C_{12}(1,0)$.

\begin{table}[h]
{\begin{tabular}{c|ccc}
\hline
{Order of neighbour} & $\mathcal{C}=C_8(1,0)$ & 
$\mathcal{C}=C_{10}(1,0)$ & $\mathcal{C}=C_{12}(1,0)$\\
\hline
0  & 0.0000 & 0.0000 & 0.0000 \\
1  & 0.1213 & 0.1755 & 0.2679 \\
2  & 0.1715 & 0.2839 & 0.3789 \\
3  & 0.2241 & 0.3338 & 0.4640 \\
4  & 0.2928 & 0.4382 & 0.5176 \\
5  & 0.3170 & 0.4566 & 0.5883 \\
6  & 0.3394 & 0.4595 & 0.5977 \\
7  & 0.3882 & 0.4891 & 0.6225 \\
8  & 0.4142 & 0.5082 & 0.6415 \\
9  & 0.4316 & 0.5377 & 0.6550 \\
10 & 0.4483 & 0.5401 & 0.6859 \\ 
\hline
\end{tabular}}
\caption{Distances to the space $E$ of the nearest points of $\mathbb{Z}^k$ in the case of
octagonal, decagonal and dodecagonal clusters used in figures 1 and 2.}
\end{table}

The left hand side of figure 3 suggests us that in the case of the sets $\mathcal{P}$ 
defined by multi-shell
$G$-clusters there is no cluster occuring at a high frequency. Therefore, generally, the 
algorithm presented in previous section does not lead to useful models. In addition, this
purely mathematical algorithm does not provide any physical insight on why the atoms should 
favour such a structure. In this section our aim is to present a modified version which leads
to interesting quasiperiodic sets, and might be justified from a physical point of view.

We start from a $G$-cluster and embed the 'physical' space $E$ into the superspace 
$\mathbb{R}^k$ in the same way as in the previous section. Then we translate $E$ by a certain
vector $t\in \mathbb{R}^k$ and consider the set 
\[ \mathcal{L}=\{ x\in \mathbb{Z}^k \ |\ \ ||x-t||<R\ \} \]
of all the points of $\mathbb{Z}^k$ lying inside the ball of centre $t$ and a fixed radius $R$. 
For each point $x\in \mathcal{L}$ we determine the distance $d(x)$ from $x$ to the space $t+E$. 
			  
\begin{figure}[b]
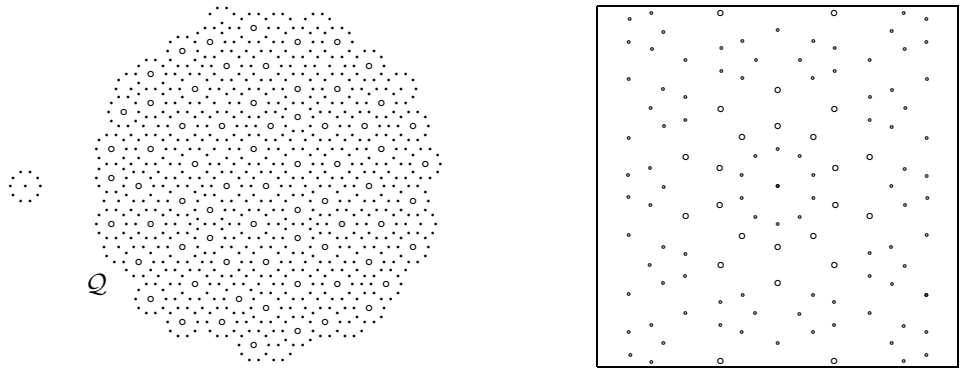
			 
\setlength{\unitlength}{2.0mm}	   
		    
\caption{{\it Left:} Quasiperiodic packing of one-shell decagonal clusters  
obtained by using the modified strip projection method. The centers of the fully 
symmetric clusters are indicated by $\circ $. {\it Right:} 
The simulated diffraction pattern.}  
\end{figure}

\begin{figure}
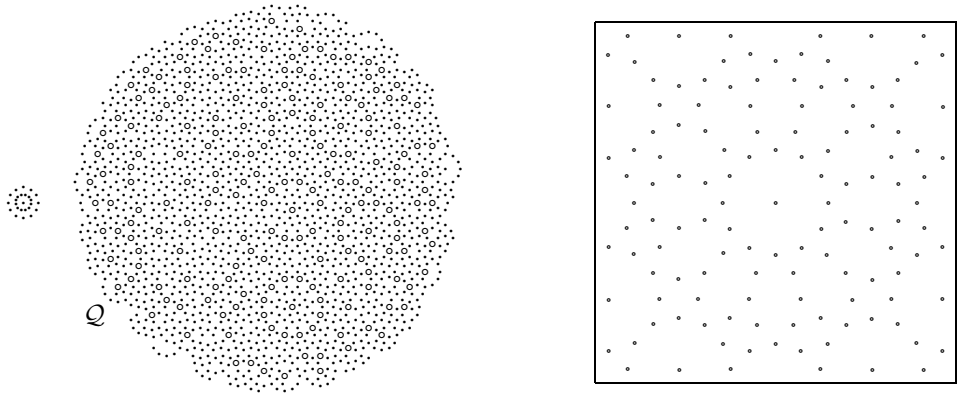
			 
\setlength{\unitlength}{2.0mm}	   
%
		   
 \caption{{\it Left:} Quasiperiodic packing of two-shell dodecagonal clusters  obtained by 
using the modified projection method and the one-shell $D_{12}$-cluster $\mathcal{C}=D_{12}(1,0)$
as starting cluster. The centers of the clusters with outer shell fully occupied are indicated
by $\circ $. {\it Right:} The simulated diffraction pattern contains only extremely sharp 
Bragg peaks, disposed acording to the $D_{12}$ symmetry. }  
\end{figure}

We define a subset $\mathcal{Q}\subset \pi (\mathbb{Z}^k)$ by starting from the assumptions:
\begin{itemize}
\item[-] the distance between any two points of $\mathcal{Q}$ is greater than a fixed minimal
distance $\delta $
\item[-] each point $q=\pi x$ from $\mathcal{Q}$ has the tendency to become the center of 
a translated copy of $\mathcal{C}$, and this tendency diminishes when $d(x)$ increases.
\end{itemize}
We generate a fragment of $\mathcal{Q}$ by considering the points $x$ of $\mathcal{L}$ 
in the increasing order of $d(x)$. 
If the distance from $\pi x$ to one of the already obtained points of $\mathcal{Q}$ is less
than $\delta $ then we pass to the next point of $\mathcal{L}$. 
If the distance from $\pi x$ to the already obtained points of $\mathcal{Q}$ 
is greater than $\delta $ then we add $\pi x$ to $\mathcal{Q}$ and continue by adding all
the points of the cluster $\pi x+\mathcal{C}$ satisfying  the same restriction (distance from
the considered point to the already obtained points of $\mathcal{Q}$ 
is greater than $\delta $). Then we pass to the next point of $\mathcal{L}$. After a certain number of steps the retriction concerning the minimal distance does not allow us to add any 
point. 

The set $\mathcal{Q}$ from figure 4, obtained by starting from the cluster 
$\mathcal{C}=D_{10}(1,0)$, contains a significant percentage of fully symmetric clusters,
and the simulated diffraction pattern consists in sharp Bragg peaks disposed according to 
$D_{10}$ symmetry. The minimal distance imposed in this case is the distance between two
neighbouring points of $\mathcal{C}$.
The remarkable set presented in figure 5 is obtained by starting from the cluster 
$\mathcal{C}=D_{12}(1,0)$ by imposing also $\delta $ to be the distance between neighbouring points of $\mathcal{C}$. The starting cluster is a one-shell cluster but in $\mathcal{Q}$ 
we found a packing of two-shell $D_{12}$-clusters with the outer shell fully occupied and 
the inner shell containing only 5-6 points. The simulated diffraction pattern corresponding 
to $\mathcal{Q}$ contains only extremly sharp Bragg peaks and admits $D_{12}$ as a symmetry group.
The computer programmes used in this paper are available on line \cite{c}.

\end{document}